\documentclass[aps,prl,notitlepage,reprint,superscriptaddress]{revtex4-1}

\usepackage[colorlinks,linkcolor=red,citecolor=blue,urlcolor=red]{hyperref}

\usepackage{float}
\usepackage{CJK}
\usepackage{graphicx}%,float}
\usepackage{amsmath,amssymb,amsfonts}

\usepackage{siunitx}
\usepackage{float}
\usepackage{mathtools}
\usepackage{soul}
\usepackage{changes}

\usepackage{amsmath,amssymb,amsfonts} % standard AMS packages
\usepackage{bm}	% bold symbols in math mode \bm{...}
\renewcommand{\mathbf}{\bm}

\usepackage{dsfont}	% proper mathbb format
\renewcommand{\mathbb}{\mathds}	% redefine \mathbb

\usepackage{mathrsfs} % use \mathscr{} for script letters in math
\usepackage{mathtools} % for proper typesetting of := and =:
\usepackage{natbib}

\renewcommand{\eqref}[1]{Eq.~(\ref{#1})}

\begin{document}

\title{Thermo-refractive noise in silicon nitride microresonators}

\author{Guanhao Huang}
\affiliation{Institute of Physics (IPHYS), {\'E}cole Polytechnique F{\'e}d{\'e}rale de Lausanne, 1015 Lausanne, Switzerland}

\author{Erwan Lucas}
\affiliation{Institute of Physics (IPHYS), {\'E}cole Polytechnique F{\'e}d{\'e}rale de Lausanne, 1015 Lausanne, Switzerland}

\author{Junqiu Liu}
\affiliation{Institute of Physics (IPHYS), {\'E}cole Polytechnique F{\'e}d{\'e}rale de Lausanne, 1015
		Lausanne, Switzerland}
	
\author{Arslan S. Raja}
\affiliation{Institute of Physics (IPHYS), {\'E}cole Polytechnique F{\'e}d{\'e}rale de Lausanne, 1015 Lausanne, Switzerland}

\author{Grigory Lihachev}
\affiliation{Faculty of Physics, M.V. Lomonosov Moscow State University, Moscow 119991, Russia}
\affiliation{Russian Quantum Centre, 143025 Skolkovo, Russia}

\author{Michael L. Gorodetsky}
\affiliation{Faculty of Physics, M.V. Lomonosov Moscow State University, Moscow 119991, Russia}
\affiliation{Russian Quantum Centre, 143025 Skolkovo, Russia}

\author{Nils J. Engelsen}
\email{nils.engelsen@epfl.ch}
\affiliation{Institute of Physics (IPHYS), {\'E}cole Polytechnique F{\'e}d{\'e}rale de Lausanne, 1015 Lausanne, Switzerland}

\author{Tobias J. Kippenberg}
\email{tobias.kippenberg@epfl.ch}
\affiliation{Institute of Physics (IPHYS), {\'E}cole Polytechnique F{\'e}d{\'e}rale de Lausanne, 1015 Lausanne, Switzerland}
%\email{tobias.kippenberg@epfl.ch}

\date{\today}

\begin{abstract}
Thermodynamic noise places a fundamental limit on the frequency stability of dielectric optical resonators. Here, we present the characterization of thermo-refractive noise in photonic-chip-based silicon nitride ($\text{Si}_3\text{N}_4$) microresonators and show that thermo-refractive noise is the dominant thermal noise source in the platform. We employed balanced homodyne detection to measure the thermo-refractive noise spectrum of microresonators of different diameters. The measurements are in good agreement with theoretical models and finite element method (FEM) simulations. Our characterization sets quantitative bounds on the scaling and absolute magnitude of thermal noise in photonic chip-based microresonators. An improved understanding of thermo-refractive noise can prove valuable in the design considerations and performance limitations of future photonic integrated devices.
\end{abstract}
\maketitle

\section{Introduction}\label{sec:intro}

Optical microresonators are used in numerous practical applications, in particular integrated, silicon-based lasers \cite{liang2010recent}, narrow linewidth lasers \cite{gundavarapu2019sub} and photonic microwave oscillators \cite{capmany2007microwave}. The small mode volume of microresonators enables strong optical nonlinearities harnessed in emerging technologies such as microcombs \cite{kippenberg2018dissipative}. The small mode volume also enhances sensing capabilities used in fundamental research, e.g. cavity quantum optomechanics \cite{aspelmeyer2014cavity}. However, the small mode volume comes at a cost: limitations on the microresonator's frequency stability arise from thermal fluctuations such as thermo-refractive (TRN) and thermoelastic noise. These fluctuations were first theoretically described in the context of Laser Interferometer Gravitational-Wave Detectors \cite{braginsky2000thermo}, and place limits on the frequency stability of an interferometer. Thermal fluctuations are particularly strong in small mode volume optical resonators and place fundamental limits on applications that require high frequency stability, e.g. optical sensing \cite{Foreman:15}, optomechanical displacement sensing \cite{AnetsbergerPRA}, dissipative Kerr soliton microcomb generation \cite{kippenberg2018dissipative}, electro-optical modulators \cite{Pavlov:15}, opto-electronic oscillators \cite{liang2015high} and Kerr squeezing \cite{hoff2015integrated}. Therefore, different kinds of thermal noises have been extensively studied \cite{braginsky2000thermo,gorodetsky2004fundamental,Matsko:07,weng2018ultra}, but theoretical analyses show inconsistencies between different platforms \cite{kondratiev2018thermorefractive} and rely on auxiliary measurements of material parameters that are not always well-known. Experimental characterization is therefore essential to understand the limitations of a specific system. %Here, we present a method to characterize the thermo-refractive noise of microresonators and apply it to $\text{Si}_3\text{N}_4$ microresonators.

Silicon nitride ($\text{Si}_3\text{N}_4$) is a space compatible \cite{brasch2014radiation}, CMOS-compatible material \cite{moss2013new} with a large Kerr nonlinear coefficient, an absence of two photon absorption in the telecommunication window, ultralow losses \cite{pfeiffer2016photonic,ji2017ultra}, and a wide transparency window from visible to mid-infrared. 
These properties have in particular been exploited for chip-scale frequency combs \cite{kippenberg2011microresonator,kippenberg2018dissipative}, as well as coherent low-pulse-energy supercontinuum generation in the near- \cite{johnson2015octave} and mid-infrared \cite{guo2018mid}.
%It is therefore widely used to fabricate integrated waveguides and high-Q microresonators. Using $\text{Si}_3\text{N}_4$ microresonators, soliton-based microresonator frequency combs \cite{kippenberg2011microresonator,kippenberg2018dissipative} can be generated with FSR ranging from \SI{20}{GHz} to \SI{1}{THz}. 
Recent advances \cite{pfeiffer2018photonic,Liu:18,ji2017ultra} in fabrication of integrated $\text{Si}_3\text{N}_4$ microresonators have enabled optical quality factors $Q>10^7$, for which the fundamental limits imposed by TRN will be relevant for the frequency stability, and can be a limiting factor in future applications such as microwave generation \cite{capmany2007microwave}.
%Recent advances \cite{pfeiffer2016photonic,Liu:18,ji2017ultra} in fabrication of integrated $\text{Si}_3\text{N}_4$ microresonators have enabled optical quality factors high enough to explore the fundamental limits on the microresonator frequency stability, which will be a limiting factor in future applications such as microwave generation \cite{capmany2007microwave}. 
Although recent measurements of the carrier-envelope-offset frequency noise of microcombs \cite{drake2018thermal} have already shown clear evidence that thermal noise is limiting the performance of certain applications, TRN has so far only been investigated in simple geometries (notably silica microspheres and microtoroids \cite{AnetsbergerPRA,gorodetsky2004fundamental}), and has not yet been measured in photonic integrated microresonators.

Here, we present the first characterization of TRN in $\text{Si}_3\text{N}_4$ microresonators,  and compare FEM simulations with measurements of TRN using balanced homodyne detection. The results show that TRN is the dominant thermal noise source over frequencies ranging from \SI{10}{\kHz} to \SI{10}{\MHz}. Refined theoretical and modeling approaches are therefore required to analyze the TRN in $\text{Si}_3\text{N}_4$ microresonators. %Our results improve the  understanding of TRN in a complex photonically integrated system. [This sentence is very out of place --NJE]

\section{Thermo-Refractive Noise}
\label{sec:theory}
In an optical resonator, thermo-refractive noise leads to fluctuations of the resonance frequency due to fluctuations of refractive index, $n$, caused by thermodynamic fluctuations of temperature, $\delta T$, whose variance is:
\begin{equation}
    \langle \delta T^2\rangle=\frac{k_B T^2}{\rho C V}
\end{equation}
where $T$ is the temperature of the heat bath, $k_B$ the Boltzmann constant, $\rho$ the density, $C$ the specific heat, and $V$ the volume. Using the material parameters of $\text{Si}_3\text{N}_4$ and the optical mode volume of a typical \SI{1}{THz} free-spectral-range (FSR) microresonator, we can obtain a value of the standard deviation of the heat bath temperature as $\sqrt{\langle \delta T^2\rangle}\sim\SI{60}{\micro K}$. Combined with the measured thermo-optic coefficient, $\mathrm{d}n/\mathrm{d}T=\SI{2.45e-5}{K^{-1}}$, of $\text{Si}_3\text{N}_4$ \cite{dndt}, the fractional frequency fluctuation can be estimated as\cite{gorodetsky2004fundamental} $\sqrt{\langle \delta f^2\rangle}/f\sim7\times10^{-10}$, and the absolute frequency fluctuation $\sqrt{\langle \delta f^2\rangle}$ is around \SI{150}{kHz}, which is above $1\%$ of the cavity linewidth in $\text{Si}_3\text{N}_4$ microresonators \cite{Liu:18}.% and is comparable in value to the narrow linewidths of crystalline microresonators based on $\mathrm{MgF_2}$ and $\mathrm{CaF_2}$ \cite{Savchenkov:07}.

The thermo-refractive noise and the thermo-elastic noise, which are both consequences of thermodynamic temperature fluctuations, were previously experimentally observed in silica microspheres \cite{gorodetsky2004fundamental} and theoretically analyzed (but not observed) in crystalline resonators \cite{Matsko:07}. In most cases, thermo-refractive noise is the largest among the thermal noises. For chip-based $\text{Si}_3\text{N}_4$ microresonators, TRN is also expected to be the largest noise source, as the thermo-optic coefficient ($\mathrm{d}n/\mathrm{d}T$) is larger than the thermal expansion coefficient. However, the different modelling approaches taken in the analysis of the two previously mentioned platforms result in different predictions for the geometric dependency of TRN as well as its magnitude at low offset frequencies \cite{kondratiev2018thermorefractive}.

We now describe the two approaches taken for modelling TRN in microresonators: the first model assumes a homogeneous microresonator in an infinite heat bath, yielding the following expression for the effective temperature fluctuations: \cite{kondratiev2018thermorefractive}:
\begin{equation}\label{eq:inf}
    S_{\delta T}(\omega)=\frac{k_BT^2}{\sqrt{\pi^3\kappa\rho C\omega}}\sqrt{\frac{1}{2p+1}}\frac{1}{R\sqrt{d_z^2-d_r^2}}\frac{1}{(1+(\omega\tau_d)^{3/4})^2}
\end{equation}
where $R$ is the ring radius of the microcavity (geometry shown in Fig.~\ref{fig:results}(a)), $d_z$ and $d_r$ are halfwidths of the fundamental mode with orbital number $l$, azimuthal number $m$, meridional mode number $p=l-m$, $\tau_d=\frac{\pi^{1/3}}{4^{1/3}}\frac{\rho C}{\kappa}d_r^2$, and the definitions of the other parameters can be found in Table \ref{tab:phys}. The key features are the $\omega^{-1/2}$ dependence at low frequency, the $\omega^{-2}$ dependence at high frequency, and the $R^{-1}$ overall scaling. This model gave satisfactory agreements with experimental measurements in microspheres \cite{gorodetsky2004fundamental}. 

The second model \cite{Matsko:07} uses the thermal decomposition method which does not take into account the interaction with the environment. As a consequence, there is a low frequency cut-off due to the finite dimension of the resonator, which results in the following (single-sided) temperature noise spectral density:
\begin{equation}\label{eq:fin}
    S_{\delta T}(\omega)=\frac{k_BT^2R^2}{12\kappa V_{\mathrm{eff}}}\left(1+\left(\frac{R^2\rho C\omega}{3^{5/3}\kappa}\right)^{3/2}+\frac{1}{6}\left(\frac{R^2\rho C\omega}{8l^{1/3}\kappa}\right)^{2}\right)^{-1}
\end{equation}
where $V_{\mathrm{eff}}$ is the effective mode volume. Here the expression still has a $\omega^{-2}$ dependence at high frequency, but features an overall scaling with $R$ between $R^{-3}$ and $R^{-4}$ depending on how $V_{\mathrm{eff}}$ scales with radius.

The temperature fluctuations can be converted into a frequency noise spectral density through $S_{\delta f} = (f_0\frac{1}{n_0} \frac{dn}{dT})^2 S_{\delta T}$, where $f_0$ is the resonance frequency. However,  Eq.~(\ref{eq:inf}) and Eq.~(\ref{eq:fin}) are both idealized cases assuming homogeneous materials and either the infinite heat bath or isolated model. Clearly, these assumptions do not match the geometry of integrated $\text{Si}_3\text{N}_4$ resonators consisting of complex waveguide structures comprising different materials, which invalidates the assumption of homogeneity. We therefore performed an FEM simulation based on the fluctuation-dissipation theorem \cite{kondratiev2018thermorefractive,weng2018ultra}, where we simulate the thermo-refractive noise of a $\text{Si}_3\text{N}_4$ microresonator embedded in a $\text{Si}\text{O}_2$ substrate, using the actual geometry.

\begin{table}[h]
\centering
\caption{\bf Physical properties used for both the theoretical models and FEM simulations of the thermo-refractive noise of $\text{Si}_3\text{N}_4$ micoresonators}
\begin{tabular}{|c|c|}
\hline
Physical properties & Values\\
\hline
Density ($\rho$) & \SI{3.29e3}{kg.m^{-3}}\\
Refractive index ($n_0$) & 1.996\\
Thermo-optic coefficient ($\mathrm{d}n/\mathrm{d}T$) & \SI{2.45e-5}{K^{-1}}\\
Thermal conductivity ($\kappa$) & \SI{30}{W.m^{-1}.K^{-1}}\\
Specific heat capacity ($C$) & \SI{800}{J.kg^{-1}.K^{-1}}\\
\hline
\end{tabular}
  \label{tab:phys}
\end{table}

\begin{figure}[h]
\centering
\includegraphics[width=\linewidth]{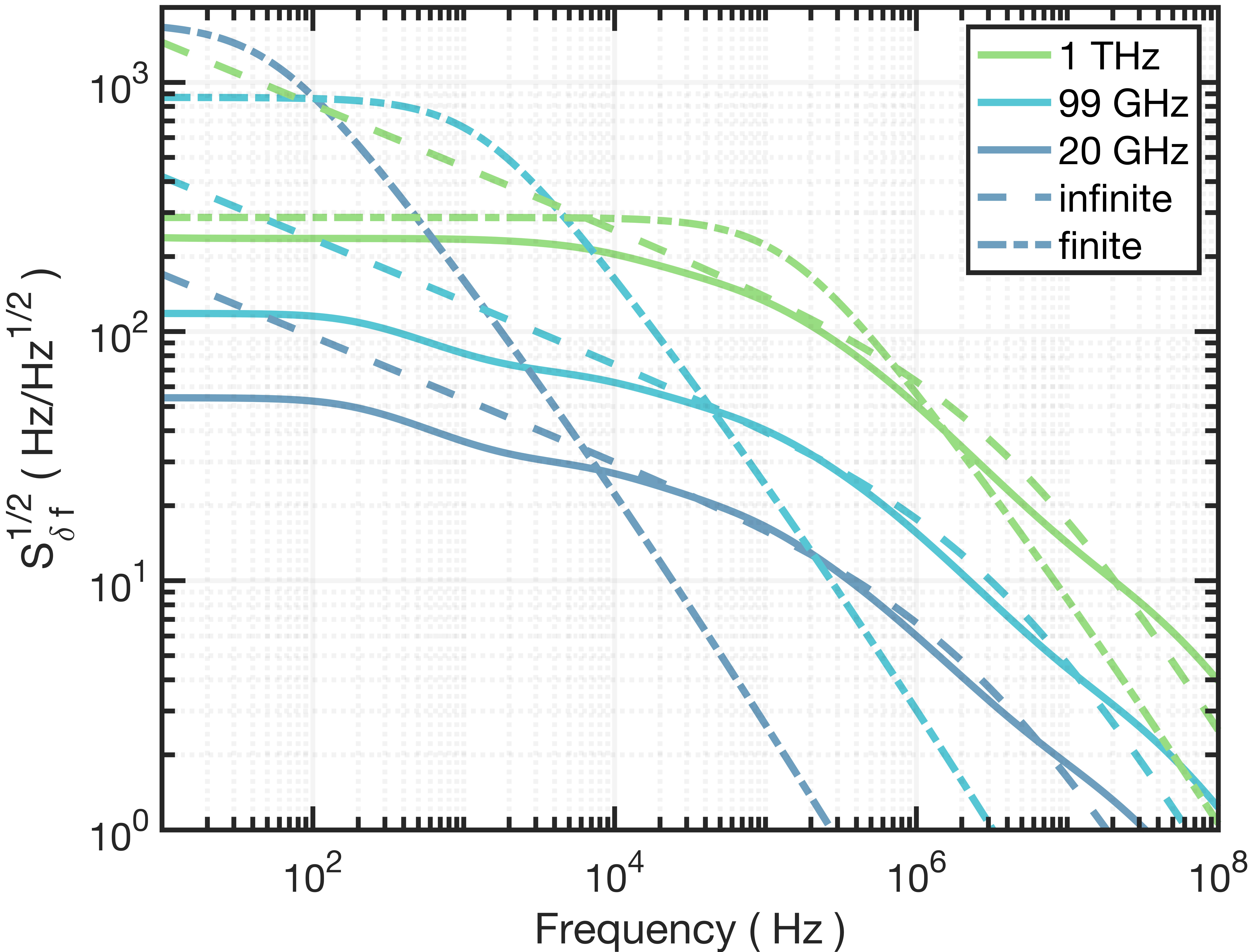}
\caption{\textbf{Comparison of FEM simulation results of thermo-refractive noise with the theoretical predictions (infinite heat bath and thermal decomposition)}. The graphs show the cavity frequency noise spectral density of a $\text{Si}_3\text{N}_4$ microresonator with a free spectral range (FSR) of \SI{1}{THz}, \SI{99}{GHz} and \SI{20}{GHz}. At high frequency, the simulation curves match better with the infinite heat bath assumption. At low frequency, the simulation curves experience a cut-off because of the finite size of the modelled chip, which behaves more similar to the thermal decomposition method.}
\label{fig:comp}
\end{figure}

We can now compare the FEM simulation results with the theoretical expressions (see Fig.~\ref{fig:comp}). Because the thermal properties of $\text{Si}_3\text{N}_4$ depend on the material characteristics and the fabrication process, they can exhibit significant variation. We use the median values of the physical properties reported in the literature for both the theoretical predictions and the FEM simulations (see Table \ref{tab:phys}). The optical mode parameters used in Eq.~(\ref{eq:inf}) and Eq.~(\ref{eq:fin}) are retrieved from FEM simulations of the $\text{Si}_3\text{N}_4$ microresonators. For the scaling with ring radius, the simulation curves match the infinite heat bath theory Eq.~(\ref{eq:inf}) well, while the deviation from the thermal decomposition method becomes larger as the radius increases. It indicates that, according to FEM simulations, $\text{Si}_3\text{N}_4$ microresonators experience thermo-refractive noise more similar to microspheres at high frequency, while at low frequency, the noise is reduced due to the cladding region ($\text{Si}\text{O}_2$) and the chip geometry. The measurement results, the simulation curves and the infinite heat bath curves are compared in Fig.~\ref{fig:results}.  

Due to the $R^{-1}$ scaling and the strong light confinement the $\text{Si}_3\text{N}_4$ waveguide offers, the computed spectral density of thermo-refractive frequency noise for resonators with FSR ranging from 20~GHz and 1~THz is sufficiently high to be probed without an extremely pure and stable laser. 
In crystalline microresonators, the much larger mode volume and smaller thermo-optic coefficient ($\mathrm{d}n/\mathrm{d}T$) make it much more challenging to measure TRN~\cite{Lim2017a}, and correspondingly it is typically not a practical limit.

\begin{figure}[h]
\centering
\includegraphics[width=\linewidth]{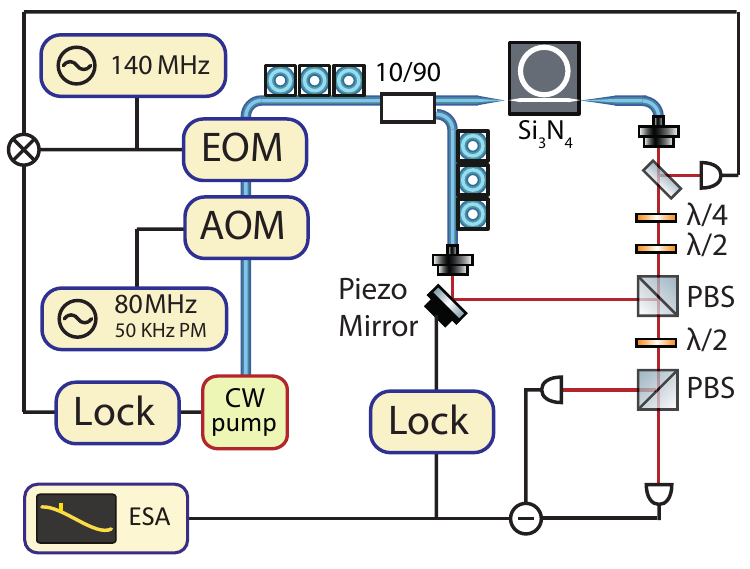}
\caption{\textbf{Balanced homodyne setup for measuring thermo-refractive noise of the $\text{Si}_3\text{N}_4$ microresonator.} The electro-optic modulator (EOM) phase-modulates the light at \SI{140}{\MHz} to generate the PDH error signal for locking the laser to the cavity mode. The AOM is modulated with an RF tone with a central frequency of \SI{80}{MHz}. The RF tone is then phase modulated at \SI{50}{kHz} with a \SI{1.14}{rad} modulation depth, which provides an absolute calibration peak for the noise measurements. A piezo mirror is used in the local oscillator path to lock the homodyne at the phase quadrature. The measurement bandwidth is limited by the detector bandwidth of \SI{80}{MHz}. }
\label{fig:setup}
\end{figure}

\section{Measurement scheme and results}
%\subsection{Measurement scheme}
The measurement scheme in this study employs a balanced homodyne setup to measure the phase fluctuation of the transmitted laser caused by the cavity frequency noise. The cavity frequency noise is calibrated using a calibration tone. The laser (external cavity diode laser at $\SI{1550}{nm}$, linewidth $\sim 30$~kHz) is locked to the cavity resonance via the Pound-Drever-Hall (PDH) locking method \cite{drever1983laser} with a feedback bandwidth of a few kilohertz. The power level is sufficiently low to avoid any unwanted thermal effects. The measurement setup is illustrated in Fig.~\ref{fig:setup}.

In order to provide an absolute calibration of our measurement, we use a calibration tone generated by phase modulating the RF tone applied to the acousto-optic modulator (AOM). A modulation depth of \SI{1.14}{rad} is calibrated by sideband fitting using heterodyne detection. The calibration tone induces an extra phase fluctuation transduced by the cavity, and results in a narrow peak in the frequency noise spectrum. Because the frequency modulation depth is known for the calibration peak, we can use it to calibrate the absolute magnitude of the corresponding homodyne signal. The modulation frequency and the modulation depth are chosen at \SI{50}{kHz} and \SI{1.14}{rad} to be outside the PDH locking bandwidth and to keep the frequency modulation depth smaller than the cavity linewidth.

The characterized samples are integrated $\text{Si}_3\text{N}_4$ microresonators with radius $R\sim23-\SI{1200}{\micro m}$, and FSR ranging from \SI{1}{THz} to \SI{20}{GHz}. The \textit{Q}-factors of these microresonators fabricated by the photonic Damascene reflow process \cite{Liu:18,pfeiffer2018photonic} are typically $Q>10^7$. The measured noise spectrum is thus filtered by the cavity resonance at high offset frequencies. The resonance linewidth of each microresonator and the response function of the bias tee before the spectrum analyzer are measured and compensated for through data post-processing.

%\begin{figure}[hb]
%\centering
%\includegraphics[width=\linewidth]{20181217_powersweep.png}
%\caption{\textbf{The thermo-refractive noise spectrums measured using different probing power.} A $\text{Si}_3\text{N}_4$ microresonator with free spectral range of \SI{1}{THz} was probed under \SI{20}{\micro W}, \SI{80}{\micro W} and \SI{120}{\micro W} of optical power, and the same level of noise spectrum is revealed. It indicates that the probe is week enough to avoid other thermal effects. It also reveals the power-independent nature of thermo-refractive noise.}
%\label{fig:powersweep}
%\end{figure}

%\subsection{Measurement results}

\begin{figure}[t]
\centering
\includegraphics[width=\linewidth]{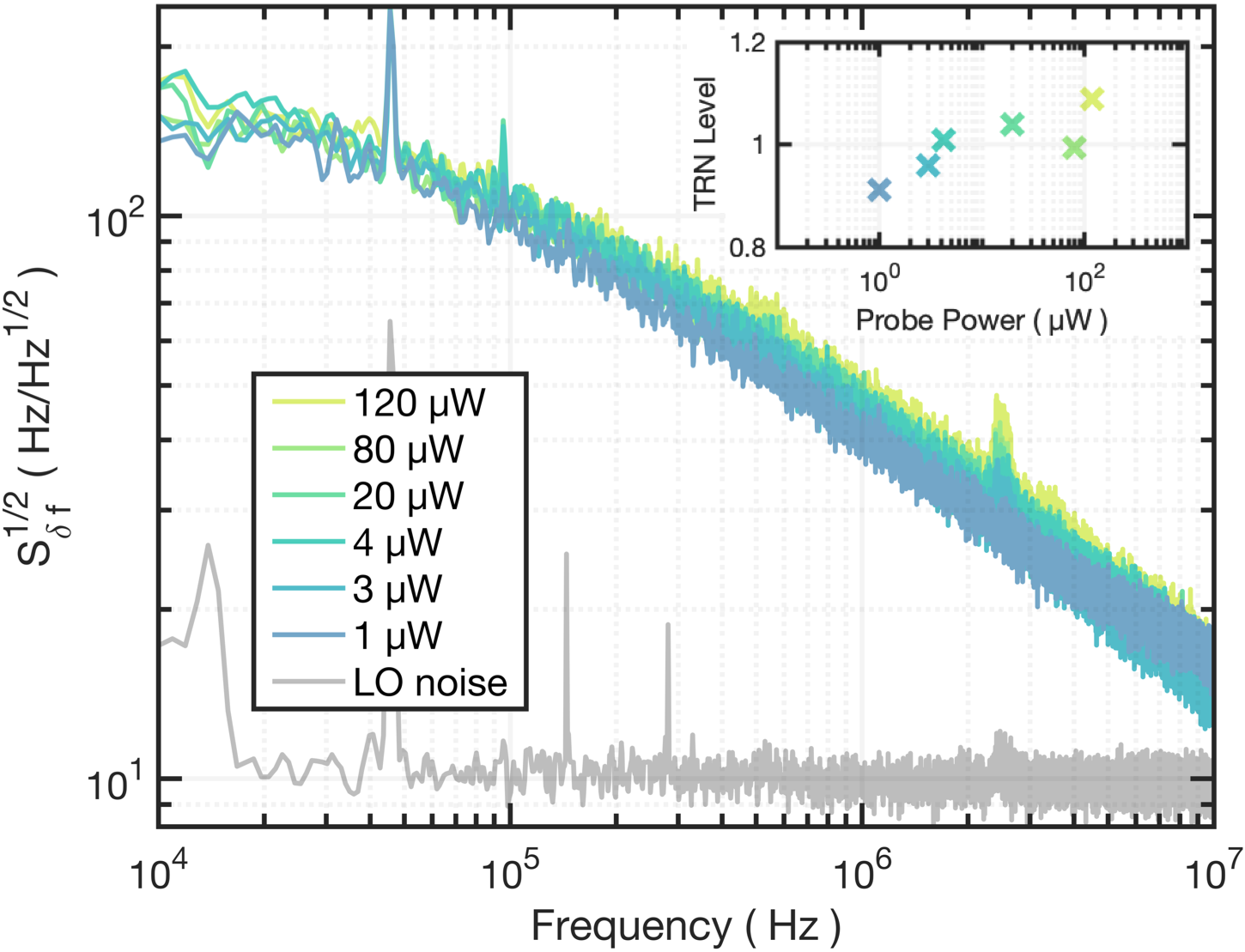}
%fbox{}
\caption{\textbf{Calibrated thermo-refractive frequency noise spectra measured using different probing power.} A $\text{Si}_3\text{N}_4$ microresonator with free spectral range of \SI{1}{THz} was probed with \SI{1}{\micro W} to \SI{120}{\micro W} of optical power. The lowest curve (grey) shows the shot noise level with \SI{1}{\micro W} of input optical power. The normalized units utilized in the inset figure are obtained by integrating over \SI{200}{kHz} to \SI{2}{MHz} and dividing by the average of the integrated values. It indicates that the probe is weak enough to avoid other laser-induced thermal effects (e.g. photothermal noise), and importantly reveals the power-independence expected for thermo-refractive noise. } %\hl{TJK: I suggest that the reference noise curve is made grey and also referenced in the legend or figure caption. In addition the inset figure has crosses that are too small to see. Use larger filled circles.}} 
\label{fig:pwrsweep}
\end{figure}

\begin{figure}[t]
\centering
\includegraphics[width=\linewidth]{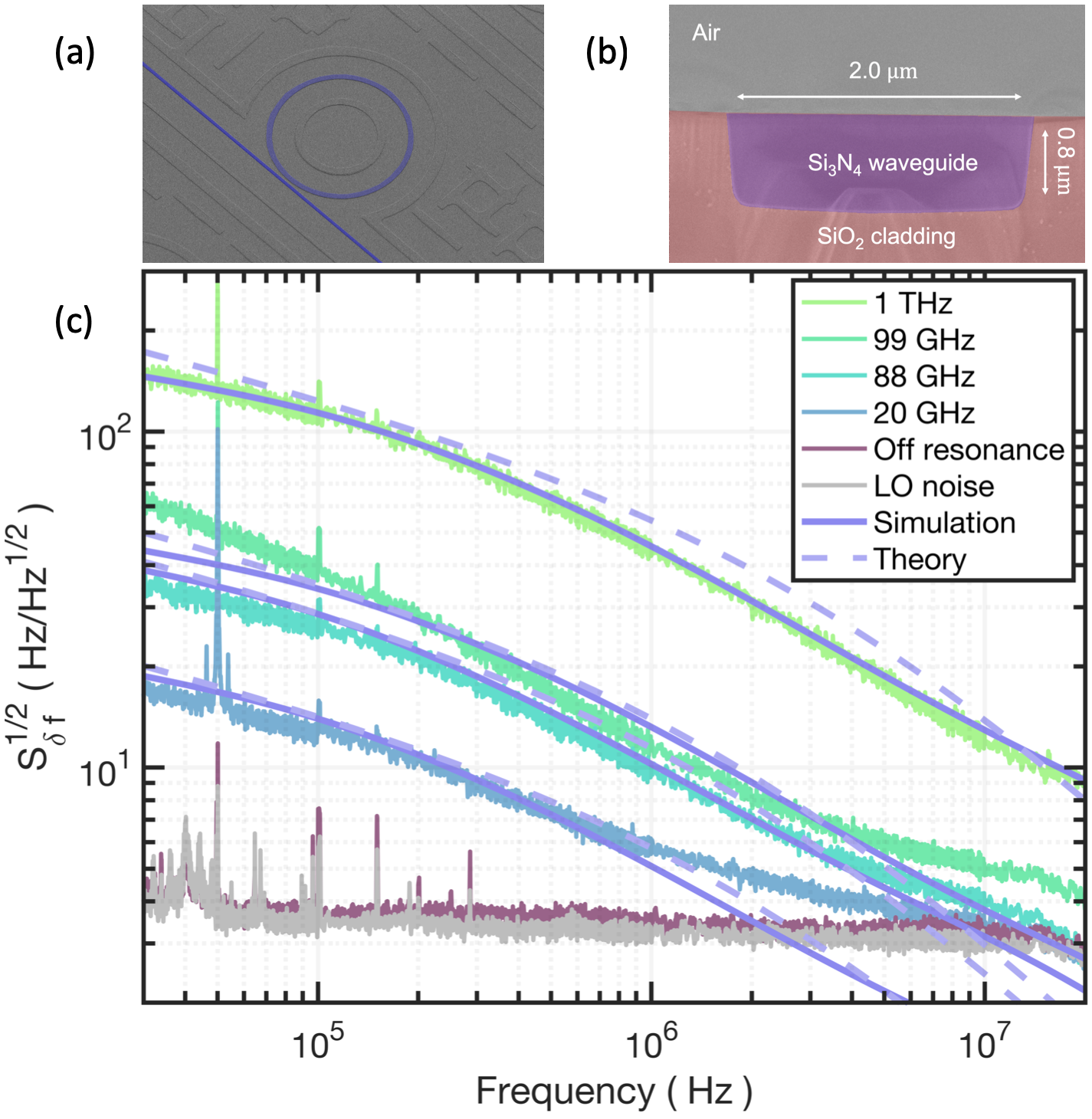}
%fbox{}
\caption{\textbf{Verifying the size dependence of thermo-refractive noise in integrated $\text{Si}_3\text{N}_4$ microresonators.} \textbf{(a)} SEM image of a \SI{1}{THz}-FSR $\text{Si}_3\text{N}_4$ microresonator ring (blue). \textbf{(b)} SEM image of the cross-section of the waveguide. The cross-section image is color shaded to help identify different regions. 
%The \SI{1}{THz} sample ring we use for TRN measurement has a smaller width of \SI{1.35}{\micro m} than the one on the image, with \SI{0.8}{\micro m} height, and a ring radius of \SI{22.75}{\micro m}.
\textbf{(c)} Thermo-refractive noise measured in integrated $\text{Si}_3\text{N}_4$ microresonators with free spectral ranges of \SI{1}{THz}, \SI{99}{GHz}, \SI{88}{GHz} and \SI{20}{GHz}. A 30-point moving average was applied to the data. The \SI{20}{GHz} off-resonance noise spectrum (brown) and the LO shot noise spectrum (i.e. homodyne signal of the LO beating with vacuum noise of the blocked signal arm, shown as gray) are also plotted and shown to overlap, indicating that the measurement is quantum-noise limited at frequencies above \SI{30}{\kHz}. The solid blue lines show the FEM simulation results and the dashed blue lines show the theoretical predictions from Eq.~(\ref{eq:inf}). Data below $\SI{10}{kHz}$ were truncated due to the excess locking noises.}%filtering used before the ESA.}%\hl{Comments: A figure caption was missing. In addition the graph would be more clear if we only show the averages, instead of the non-averaged data. Use grey for the LO shot noise. Finally, make the panels A and B the same width as the horizontal axis of panel C. }}
\label{fig:results}
\end{figure}

We first verified the power-independent nature of thermo-refractive noise (as expected from Eqs.1-3) by performing an input power sweep (shown in Fig.~\ref{fig:pwrsweep}). The frequency noise level remains the same when varying the laser power of the probe signal by more than two orders of magnitude (from $\SI{1}{\micro W}$ to $\SI{120}{\micro W} $), showing that photothermal noise is not making a significant contribution.
We next investigated the dependence on optical mode volume. Fig.~\ref{fig:results} (c) presents the measurement results for four different cavity radii, together with the corresponding theoretical curves and FEM simulation curves. The observable background noise sources arise from local oscillator shot noise, several technical spikes, and the calibration peak at \SI{50}{kHz}. The calibration peaks in the off-resonance noise spectrum and the LO shot noise spectrum from Fig.~\ref{fig:results} (c) are the result of residual amplitude modulation from the AOM. However, by utilizing phase modulation of the AOM RF tone, a signal-to-noise ratio of \SI{20}{dB} can be obtained for the calibration peak.  %\hl{We should add one sentence on the magnitude of the TRN noise, i.e. for the 1THz resonator calculate the rms cavity frequency fluctuations. It should be a sizable amount of noise.}

Good agreement of the measured spectra with the simulation curves for both frequency dependence ($\propto \omega^{-1/2}$) and radius dependence ($\propto R^{-1}$) is clearly observed, which also confirms the validity of  Eq.~(\ref{eq:inf}) as a theoretical prediction of TRN in the $\text{Si}_3\text{N}_4$ microresonator platform. By assuming that the spectrum matches the FEM simulation in the low frequency range, the total frequency fluctuations due to TRN could be retrieved through integration over the high frequency data and the low frequency FEM curves, e.g. the \SI{1}{THz} microresonator has a resonance frequency instability of around \SI{240}{kHz}, which agrees well with our previous estimate of \SI{150}{kHz}. The agreement further indicates that the heat exchange with the surrounding environment is largely responsible for the generation of thermo-refractive noise in this system. However, in the low frequency range, the approximation of a homogeneous medium in Eq.~(\ref{eq:inf}) will break down due to heat exchange with the media outside the waveguide, as is indicated by the multiple saturation steps of the FEM simulation curves at low frequency shown in Fig.~\ref{fig:comp}. Though the heat exchange with the outer layer makes it difficult to derive a simple expression for the thermo-refractive noise in $\text{Si}_3\text{N}_4$ microresonators, it offers a possibility to bypass the thermal limit \cite{noisereduction}. 
%Having a good confidence in our FEM modeling, we also examined the possibility of thermo-refractive noise reduction using FEM simulation by placing a layer of material with negative thermal-optical coefficient right above the waveguide. We observed a weak reduction of thermo-refractive noise at low frequency, while at high frequency, a noise increment due to the weak correlation between thermal fluctuations. 
The design of such a thermal-noise-reduced photonic microresonator will be more important when more applications truly reach the thermal limit of their performance, e.g. to realize integrated ultra narrow linewidth lasers.
%The data below $30\si{kHz}$ and above $30\si{MHz}$ are not shown in Figure \ref{fig:results}. The spectrums below $30\si{kHz}$ are overwhelmed by the locking noises and are also tilted by the high pass filtering of $15\si{kHz}$, while above $30\si{MHz}$, the spectrums hit the detector bandwidth.

\section{Conclusion}

We present the first characterization of thermo-refractive noise in the integrated $\text{Si}_3\text{N}_4$ microresonator platform. The presence of thermo-refractive noise in photonic integrated resonators could limit the performance for many applications. Therefore, we measured the thermo-refractive noises of $\text{Si}_3\text{N}_4$  microresonators with a wide range of FSR. The measurement results are in good agreement with both the FEM simulation and the theoretical prediction for both frequency dependence ($\propto \omega^{-1/2}$) and radius dependence ($\propto R^{-1}$). The characterization of thermo-refractive noise in $\text{Si}_3\text{N}_4$ microresonator systems can serve as a standard for comparison of the noise features observed in associated applications such as microwave generation. Finally, these results might provide new insights for the silicon photonics community, enabling a better understanding of the formation of thermo-refractive noise as well as a path towards systems which bypass thermal noise limitations.

\section*{Funding Information}
This publication was supported by Contract HR0011-15-C-0055 and HR0011181003 from the Defense Advanced Research Projects Agency (DARPA), Defense Sciences Office (DSO), by the Swiss National Science Foundation under grant agreement No.163864, and the Russian Foundation for Basic Research project 17-02-00522. 
\section*{Acknowledgments}
All samples were fabricated at the Center for MicroNanoTechnology (CMi) at EPFL. We acknowledge Rui Ning Wang for the assistance in sample fabrication.

%\section{Data availability statement}
%Data and data analysis code will be made available through \texttt{Zenodo} upon publication. All other data needed to evaluate the conclusions in the paper are present in the paper or the supplementary materials.

\bibliographystyle{apsrev4-1}
\bibliography{ref1}

\begin{thebibliography}{29}%
\makeatletter
\providecommand \@ifxundefined [1]{%
 \@ifx{#1\undefined}
}%
\providecommand \@ifnum [1]{%
 \ifnum #1\expandafter \@firstoftwo
 \else \expandafter \@secondoftwo
 \fi
}%
\providecommand \@ifx [1]{%
 \ifx #1\expandafter \@firstoftwo
 \else \expandafter \@secondoftwo
 \fi
}%
\providecommand \natexlab [1]{#1}%
\providecommand \enquote  [1]{``#1''}%
\providecommand \bibnamefont  [1]{#1}%
\providecommand \bibfnamefont [1]{#1}%
\providecommand \citenamefont [1]{#1}%
\providecommand \href@noop [0]{\@secondoftwo}%
\providecommand \href [0]{\begingroup \@sanitize@url \@href}%
\providecommand \@href[1]{\@@startlink{#1}\@@href}%
\providecommand \@@href[1]{\endgroup#1\@@endlink}%
\providecommand \@sanitize@url [0]{\catcode `\\12\catcode `\$12\catcode
  `\&12\catcode `\#12\catcode `\^12\catcode `\_12\catcode `\%12\relax}%
\providecommand \@@startlink[1]{}%
\providecommand \@@endlink[0]{}%
\providecommand \url  [0]{\begingroup\@sanitize@url \@url }%
\providecommand \@url [1]{\endgroup\@href {#1}{\urlprefix }}%
\providecommand \urlprefix  [0]{URL }%
\providecommand \Eprint [0]{\href }%
\providecommand \doibase [0]{http://dx.doi.org/}%
\providecommand \selectlanguage [0]{\@gobble}%
\providecommand \bibinfo  [0]{\@secondoftwo}%
\providecommand \bibfield  [0]{\@secondoftwo}%
\providecommand \translation [1]{[#1]}%
\providecommand \BibitemOpen [0]{}%
\providecommand \bibitemStop [0]{}%
\providecommand \bibitemNoStop [0]{.\EOS\space}%
\providecommand \EOS [0]{\spacefactor3000\relax}%
\providecommand \BibitemShut  [1]{\csname bibitem#1\endcsname}%
\let\auto@bib@innerbib\@empty
%</preamble>
\bibitem [{\citenamefont {Liang}\ and\ \citenamefont
  {Bowers}(2010)}]{liang2010recent}%
  \BibitemOpen
  \bibfield  {author} {\bibinfo {author} {\bibfnamefont {D.}~\bibnamefont
  {Liang}}\ and\ \bibinfo {author} {\bibfnamefont {J.~E.}\ \bibnamefont
  {Bowers}},\ }\href {\doibase 10.1038/nphoton.2010.167} {\bibfield  {journal}
  {\bibinfo  {journal} {Nature Photonics}\ }\textbf {\bibinfo {volume} {4}},\
  \bibinfo {pages} {511} (\bibinfo {year} {2010})}\BibitemShut {NoStop}%
\bibitem [{\citenamefont {Gundavarapu}\ \emph {et~al.}(2019)\citenamefont
  {Gundavarapu}, \citenamefont {Brodnik}, \citenamefont {Puckett},
  \citenamefont {Huffman}, \citenamefont {Bose}, \citenamefont {Behunin},
  \citenamefont {Wu}, \citenamefont {Qiu}, \citenamefont {Pinho}, \citenamefont
  {Chauhan} \emph {et~al.}}]{gundavarapu2019sub}%
  \BibitemOpen
  \bibfield  {author} {\bibinfo {author} {\bibfnamefont {S.}~\bibnamefont
  {Gundavarapu}}, \bibinfo {author} {\bibfnamefont {G.~M.}\ \bibnamefont
  {Brodnik}}, \bibinfo {author} {\bibfnamefont {M.}~\bibnamefont {Puckett}},
  \bibinfo {author} {\bibfnamefont {T.}~\bibnamefont {Huffman}}, \bibinfo
  {author} {\bibfnamefont {D.}~\bibnamefont {Bose}}, \bibinfo {author}
  {\bibfnamefont {R.}~\bibnamefont {Behunin}}, \bibinfo {author} {\bibfnamefont
  {J.}~\bibnamefont {Wu}}, \bibinfo {author} {\bibfnamefont {T.}~\bibnamefont
  {Qiu}}, \bibinfo {author} {\bibfnamefont {C.}~\bibnamefont {Pinho}}, \bibinfo
  {author} {\bibfnamefont {N.}~\bibnamefont {Chauhan}},  \emph {et~al.},\
  }\href {\doibase 10.1038/s41566-018-0313-2} {\bibfield  {journal} {\bibinfo
  {journal} {Nature Photonics}\ }\textbf {\bibinfo {volume} {13}},\ \bibinfo
  {pages} {60} (\bibinfo {year} {2019})}\BibitemShut {NoStop}%
\bibitem [{\citenamefont {Capmany}\ and\ \citenamefont
  {Novak}(2007)}]{capmany2007microwave}%
  \BibitemOpen
  \bibfield  {author} {\bibinfo {author} {\bibfnamefont {J.}~\bibnamefont
  {Capmany}}\ and\ \bibinfo {author} {\bibfnamefont {D.}~\bibnamefont
  {Novak}},\ }\href {\doibase 10.1038/nphoton.2007.89} {\bibfield  {journal}
  {\bibinfo  {journal} {Nature Photonics}\ }\textbf {\bibinfo {volume} {1}},\
  \bibinfo {pages} {319} (\bibinfo {year} {2007})}\BibitemShut {NoStop}%
\bibitem [{\citenamefont {Kippenberg}\ \emph {et~al.}(2018)\citenamefont
  {Kippenberg}, \citenamefont {Gaeta}, \citenamefont {Lipson},\ and\
  \citenamefont {Gorodetsky}}]{kippenberg2018dissipative}%
  \BibitemOpen
  \bibfield  {author} {\bibinfo {author} {\bibfnamefont {T.~J.}\ \bibnamefont
  {Kippenberg}}, \bibinfo {author} {\bibfnamefont {A.~L.}\ \bibnamefont
  {Gaeta}}, \bibinfo {author} {\bibfnamefont {M.}~\bibnamefont {Lipson}}, \
  and\ \bibinfo {author} {\bibfnamefont {M.~L.}\ \bibnamefont {Gorodetsky}},\
  }\href {\doibase 10.1126/science.aan8083} {\bibfield  {journal} {\bibinfo
  {journal} {Science}\ }\textbf {\bibinfo {volume} {361}},\ \bibinfo {pages}
  {eaan8083} (\bibinfo {year} {2018})}\BibitemShut {NoStop}%
\bibitem [{\citenamefont {Aspelmeyer}\ \emph {et~al.}(2014)\citenamefont
  {Aspelmeyer}, \citenamefont {Kippenberg},\ and\ \citenamefont
  {Marquardt}}]{aspelmeyer2014cavity}%
  \BibitemOpen
  \bibfield  {author} {\bibinfo {author} {\bibfnamefont {M.}~\bibnamefont
  {Aspelmeyer}}, \bibinfo {author} {\bibfnamefont {T.~J.}\ \bibnamefont
  {Kippenberg}}, \ and\ \bibinfo {author} {\bibfnamefont {F.}~\bibnamefont
  {Marquardt}},\ }\href {\doibase 10.1103/RevModPhys.86.1391} {\bibfield
  {journal} {\bibinfo  {journal} {Reviews of Modern Physics}\ }\textbf
  {\bibinfo {volume} {86}},\ \bibinfo {pages} {1391} (\bibinfo {year}
  {2014})}\BibitemShut {NoStop}%
\bibitem [{\citenamefont {Braginsky}\ \emph {et~al.}(2000)\citenamefont
  {Braginsky}, \citenamefont {Gorodetsky},\ and\ \citenamefont
  {Vyatchanin}}]{braginsky2000thermo}%
  \BibitemOpen
  \bibfield  {author} {\bibinfo {author} {\bibfnamefont {V.}~\bibnamefont
  {Braginsky}}, \bibinfo {author} {\bibfnamefont {M.}~\bibnamefont
  {Gorodetsky}}, \ and\ \bibinfo {author} {\bibfnamefont {S.}~\bibnamefont
  {Vyatchanin}},\ }\href {\doibase 10.1016/S0375-9601(00)00389-3} {\bibfield
  {journal} {\bibinfo  {journal} {Physics Letters A}\ }\textbf {\bibinfo
  {volume} {271}},\ \bibinfo {pages} {303} (\bibinfo {year}
  {2000})}\BibitemShut {NoStop}%
\bibitem [{\citenamefont {Foreman}\ \emph {et~al.}(2015)\citenamefont
  {Foreman}, \citenamefont {Swaim},\ and\ \citenamefont
  {Vollmer}}]{Foreman:15}%
  \BibitemOpen
  \bibfield  {author} {\bibinfo {author} {\bibfnamefont {M.~R.}\ \bibnamefont
  {Foreman}}, \bibinfo {author} {\bibfnamefont {J.~D.}\ \bibnamefont {Swaim}},
  \ and\ \bibinfo {author} {\bibfnamefont {F.}~\bibnamefont {Vollmer}},\ }\href
  {\doibase 10.1364/AOP.7.000168} {\bibfield  {journal} {\bibinfo  {journal}
  {Adv. Opt. Photon.}\ }\textbf {\bibinfo {volume} {7}},\ \bibinfo {pages}
  {168} (\bibinfo {year} {2015})}\BibitemShut {NoStop}%
\bibitem [{\citenamefont {Arcizet}\ \emph {et~al.}(2009)\citenamefont
  {Arcizet}, \citenamefont {Rivi\`ere}, \citenamefont {Schliesser},
  \citenamefont {Anetsberger},\ and\ \citenamefont
  {Kippenberg}}]{AnetsbergerPRA}%
  \BibitemOpen
  \bibfield  {author} {\bibinfo {author} {\bibfnamefont {O.}~\bibnamefont
  {Arcizet}}, \bibinfo {author} {\bibfnamefont {R.}~\bibnamefont {Rivi\`ere}},
  \bibinfo {author} {\bibfnamefont {A.}~\bibnamefont {Schliesser}}, \bibinfo
  {author} {\bibfnamefont {G.}~\bibnamefont {Anetsberger}}, \ and\ \bibinfo
  {author} {\bibfnamefont {T.~J.}\ \bibnamefont {Kippenberg}},\ }\href
  {\doibase 10.1103/PhysRevA.80.021803} {\bibfield  {journal} {\bibinfo
  {journal} {Phys. Rev. A}\ }\textbf {\bibinfo {volume} {80}},\ \bibinfo
  {pages} {021803} (\bibinfo {year} {2009})}\BibitemShut {NoStop}%
\bibitem [{\citenamefont {Pavlov}\ \emph {et~al.}(2015)\citenamefont {Pavlov},
  \citenamefont {Kondratyev},\ and\ \citenamefont {Gorodetsky}}]{Pavlov:15}%
  \BibitemOpen
  \bibfield  {author} {\bibinfo {author} {\bibfnamefont {N.~G.}\ \bibnamefont
  {Pavlov}}, \bibinfo {author} {\bibfnamefont {N.~M.}\ \bibnamefont
  {Kondratyev}}, \ and\ \bibinfo {author} {\bibfnamefont {M.~L.}\ \bibnamefont
  {Gorodetsky}},\ }\href {\doibase 10.1364/AO.54.010460} {\bibfield  {journal}
  {\bibinfo  {journal} {Appl. Opt.}\ }\textbf {\bibinfo {volume} {54}},\
  \bibinfo {pages} {10460} (\bibinfo {year} {2015})}\BibitemShut {NoStop}%
\bibitem [{\citenamefont {Liang}\ \emph {et~al.}(2015)\citenamefont {Liang},
  \citenamefont {Eliyahu}, \citenamefont {Ilchenko}, \citenamefont
  {Savchenkov}, \citenamefont {Matsko}, \citenamefont {Seidel},\ and\
  \citenamefont {Maleki}}]{liang2015high}%
  \BibitemOpen
  \bibfield  {author} {\bibinfo {author} {\bibfnamefont {W.}~\bibnamefont
  {Liang}}, \bibinfo {author} {\bibfnamefont {D.}~\bibnamefont {Eliyahu}},
  \bibinfo {author} {\bibfnamefont {V.~S.}\ \bibnamefont {Ilchenko}}, \bibinfo
  {author} {\bibfnamefont {A.~A.}\ \bibnamefont {Savchenkov}}, \bibinfo
  {author} {\bibfnamefont {A.~B.}\ \bibnamefont {Matsko}}, \bibinfo {author}
  {\bibfnamefont {D.}~\bibnamefont {Seidel}}, \ and\ \bibinfo {author}
  {\bibfnamefont {L.}~\bibnamefont {Maleki}},\ }\href {\doibase
  10.1038/ncomms8957} {\bibfield  {journal} {\bibinfo  {journal} {Nature
  communications}\ }\textbf {\bibinfo {volume} {6}},\ \bibinfo {pages} {7957}
  (\bibinfo {year} {2015})}\BibitemShut {NoStop}%
\bibitem [{\citenamefont {Hoff}\ \emph {et~al.}(2015)\citenamefont {Hoff},
  \citenamefont {Nielsen},\ and\ \citenamefont
  {Andersen}}]{hoff2015integrated}%
  \BibitemOpen
  \bibfield  {author} {\bibinfo {author} {\bibfnamefont {U.~B.}\ \bibnamefont
  {Hoff}}, \bibinfo {author} {\bibfnamefont {B.~M.}\ \bibnamefont {Nielsen}}, \
  and\ \bibinfo {author} {\bibfnamefont {U.~L.}\ \bibnamefont {Andersen}},\
  }\href {\doibase 10.1364/OE.23.012013} {\bibfield  {journal} {\bibinfo
  {journal} {Optics Express}\ }\textbf {\bibinfo {volume} {23}},\ \bibinfo
  {pages} {12013} (\bibinfo {year} {2015})}\BibitemShut {NoStop}%
\bibitem [{\citenamefont {Gorodetsky}\ and\ \citenamefont
  {Grudinin}(2004)}]{gorodetsky2004fundamental}%
  \BibitemOpen
  \bibfield  {author} {\bibinfo {author} {\bibfnamefont {M.~L.}\ \bibnamefont
  {Gorodetsky}}\ and\ \bibinfo {author} {\bibfnamefont {I.~S.}\ \bibnamefont
  {Grudinin}},\ }\href {\doibase 10.1364/JOSAB.21.000697} {\bibfield  {journal}
  {\bibinfo  {journal} {J. Opt. Soc. Am. B}\ }\textbf {\bibinfo {volume}
  {21}},\ \bibinfo {pages} {697} (\bibinfo {year} {2004})}\BibitemShut
  {NoStop}%
\bibitem [{\citenamefont {Matsko}\ \emph {et~al.}(2007)\citenamefont {Matsko},
  \citenamefont {Savchenkov}, \citenamefont {Yu},\ and\ \citenamefont
  {Maleki}}]{Matsko:07}%
  \BibitemOpen
  \bibfield  {author} {\bibinfo {author} {\bibfnamefont {A.~B.}\ \bibnamefont
  {Matsko}}, \bibinfo {author} {\bibfnamefont {A.~A.}\ \bibnamefont
  {Savchenkov}}, \bibinfo {author} {\bibfnamefont {N.}~\bibnamefont {Yu}}, \
  and\ \bibinfo {author} {\bibfnamefont {L.}~\bibnamefont {Maleki}},\ }\href
  {\doibase 10.1364/JOSAB.24.001324} {\bibfield  {journal} {\bibinfo  {journal}
  {J. Opt. Soc. Am. B}\ }\textbf {\bibinfo {volume} {24}},\ \bibinfo {pages}
  {1324} (\bibinfo {year} {2007})}\BibitemShut {NoStop}%
\bibitem [{\citenamefont {Weng}\ \emph {et~al.}(2018)\citenamefont {Weng},
  \citenamefont {Light},\ and\ \citenamefont {Luiten}}]{weng2018ultra}%
  \BibitemOpen
  \bibfield  {author} {\bibinfo {author} {\bibfnamefont {W.}~\bibnamefont
  {Weng}}, \bibinfo {author} {\bibfnamefont {P.~S.}\ \bibnamefont {Light}}, \
  and\ \bibinfo {author} {\bibfnamefont {A.~N.}\ \bibnamefont {Luiten}},\
  }\href {\doibase 10.1364/OL.43.001415} {\bibfield  {journal} {\bibinfo
  {journal} {Optics letters}\ }\textbf {\bibinfo {volume} {43}},\ \bibinfo
  {pages} {1415} (\bibinfo {year} {2018})}\BibitemShut {NoStop}%
\bibitem [{\citenamefont {Kondratiev}\ and\ \citenamefont
  {Gorodetsky}(2018)}]{kondratiev2018thermorefractive}%
  \BibitemOpen
  \bibfield  {author} {\bibinfo {author} {\bibfnamefont {N.}~\bibnamefont
  {Kondratiev}}\ and\ \bibinfo {author} {\bibfnamefont {M.}~\bibnamefont
  {Gorodetsky}},\ }\href {\doibase 10.1016/j.physleta.2017.04.043} {\bibfield
  {journal} {\bibinfo  {journal} {Physics Letters A}\ }\textbf {\bibinfo
  {volume} {382}},\ \bibinfo {pages} {2265} (\bibinfo {year}
  {2018})}\BibitemShut {NoStop}%
\bibitem [{\citenamefont {Brasch}\ \emph {et~al.}(2014)\citenamefont {Brasch},
  \citenamefont {Chen}, \citenamefont {Schiller},\ and\ \citenamefont
  {Kippenberg}}]{brasch2014radiation}%
  \BibitemOpen
  \bibfield  {author} {\bibinfo {author} {\bibfnamefont {V.}~\bibnamefont
  {Brasch}}, \bibinfo {author} {\bibfnamefont {Q.-F.}\ \bibnamefont {Chen}},
  \bibinfo {author} {\bibfnamefont {S.}~\bibnamefont {Schiller}}, \ and\
  \bibinfo {author} {\bibfnamefont {T.~J.}\ \bibnamefont {Kippenberg}},\ }\href
  {\doibase 10.1364/OE.22.030786} {\bibfield  {journal} {\bibinfo  {journal}
  {Optics Express}\ }\textbf {\bibinfo {volume} {22}},\ \bibinfo {pages}
  {30786} (\bibinfo {year} {2014})}\BibitemShut {NoStop}%
\bibitem [{\citenamefont {Moss}\ \emph {et~al.}(2013)\citenamefont {Moss},
  \citenamefont {Morandotti}, \citenamefont {Gaeta},\ and\ \citenamefont
  {Lipson}}]{moss2013new}%
  \BibitemOpen
  \bibfield  {author} {\bibinfo {author} {\bibfnamefont {D.~J.}\ \bibnamefont
  {Moss}}, \bibinfo {author} {\bibfnamefont {R.}~\bibnamefont {Morandotti}},
  \bibinfo {author} {\bibfnamefont {A.~L.}\ \bibnamefont {Gaeta}}, \ and\
  \bibinfo {author} {\bibfnamefont {M.}~\bibnamefont {Lipson}},\ }\href
  {\doibase 10.1038/nphoton.2013.183} {\bibfield  {journal} {\bibinfo
  {journal} {Nature Photonics}\ }\textbf {\bibinfo {volume} {7}},\ \bibinfo
  {pages} {597} (\bibinfo {year} {2013})}\BibitemShut {NoStop}%
\bibitem [{\citenamefont {Pfeiffer}\ \emph {et~al.}(2016)\citenamefont
  {Pfeiffer}, \citenamefont {Kordts}, \citenamefont {Brasch}, \citenamefont
  {Zervas}, \citenamefont {Geiselmann}, \citenamefont {Jost},\ and\
  \citenamefont {Kippenberg}}]{pfeiffer2016photonic}%
  \BibitemOpen
  \bibfield  {author} {\bibinfo {author} {\bibfnamefont {M.~H.}\ \bibnamefont
  {Pfeiffer}}, \bibinfo {author} {\bibfnamefont {A.}~\bibnamefont {Kordts}},
  \bibinfo {author} {\bibfnamefont {V.}~\bibnamefont {Brasch}}, \bibinfo
  {author} {\bibfnamefont {M.}~\bibnamefont {Zervas}}, \bibinfo {author}
  {\bibfnamefont {M.}~\bibnamefont {Geiselmann}}, \bibinfo {author}
  {\bibfnamefont {J.~D.}\ \bibnamefont {Jost}}, \ and\ \bibinfo {author}
  {\bibfnamefont {T.~J.}\ \bibnamefont {Kippenberg}},\ }\href {\doibase
  10.1364/OPTICA.3.000020} {\bibfield  {journal} {\bibinfo  {journal} {Optica}\
  }\textbf {\bibinfo {volume} {3}},\ \bibinfo {pages} {20} (\bibinfo {year}
  {2016})}\BibitemShut {NoStop}%
\bibitem [{\citenamefont {Ji}\ \emph {et~al.}(2017)\citenamefont {Ji},
  \citenamefont {Barbosa}, \citenamefont {Roberts}, \citenamefont {Dutt},
  \citenamefont {Cardenas}, \citenamefont {Okawachi}, \citenamefont {Bryant},
  \citenamefont {Gaeta},\ and\ \citenamefont {Lipson}}]{ji2017ultra}%
  \BibitemOpen
  \bibfield  {author} {\bibinfo {author} {\bibfnamefont {X.}~\bibnamefont
  {Ji}}, \bibinfo {author} {\bibfnamefont {F.~A.}\ \bibnamefont {Barbosa}},
  \bibinfo {author} {\bibfnamefont {S.~P.}\ \bibnamefont {Roberts}}, \bibinfo
  {author} {\bibfnamefont {A.}~\bibnamefont {Dutt}}, \bibinfo {author}
  {\bibfnamefont {J.}~\bibnamefont {Cardenas}}, \bibinfo {author}
  {\bibfnamefont {Y.}~\bibnamefont {Okawachi}}, \bibinfo {author}
  {\bibfnamefont {A.}~\bibnamefont {Bryant}}, \bibinfo {author} {\bibfnamefont
  {A.~L.}\ \bibnamefont {Gaeta}}, \ and\ \bibinfo {author} {\bibfnamefont
  {M.}~\bibnamefont {Lipson}},\ }\href {\doibase 10.1364/OPTICA.4.000619}
  {\bibfield  {journal} {\bibinfo  {journal} {Optica}\ }\textbf {\bibinfo
  {volume} {4}},\ \bibinfo {pages} {619} (\bibinfo {year} {2017})}\BibitemShut
  {NoStop}%
\bibitem [{\citenamefont {Kippenberg}\ \emph {et~al.}(2011)\citenamefont
  {Kippenberg}, \citenamefont {Holzwarth},\ and\ \citenamefont
  {Diddams}}]{kippenberg2011microresonator}%
  \BibitemOpen
  \bibfield  {author} {\bibinfo {author} {\bibfnamefont {T.~J.}\ \bibnamefont
  {Kippenberg}}, \bibinfo {author} {\bibfnamefont {R.}~\bibnamefont
  {Holzwarth}}, \ and\ \bibinfo {author} {\bibfnamefont {S.~A.}\ \bibnamefont
  {Diddams}},\ }\href {\doibase 10.1126/science.1193968} {\bibfield  {journal}
  {\bibinfo  {journal} {Science}\ }\textbf {\bibinfo {volume} {332}},\ \bibinfo
  {pages} {555} (\bibinfo {year} {2011})}\BibitemShut {NoStop}%
\bibitem [{\citenamefont {Johnson}\ \emph {et~al.}(2015)\citenamefont
  {Johnson}, \citenamefont {Mayer}, \citenamefont {Klenner}, \citenamefont
  {Luke}, \citenamefont {Lamb}, \citenamefont {Lamont}, \citenamefont {Joshi},
  \citenamefont {Okawachi}, \citenamefont {Wise}, \citenamefont {Lipson},\ and\
  \citenamefont {Gaeta}}]{johnson2015octave}%
  \BibitemOpen
  \bibfield  {author} {\bibinfo {author} {\bibfnamefont {A.~R.}\ \bibnamefont
  {Johnson}}, \bibinfo {author} {\bibfnamefont {A.~S.}\ \bibnamefont {Mayer}},
  \bibinfo {author} {\bibfnamefont {A.}~\bibnamefont {Klenner}}, \bibinfo
  {author} {\bibfnamefont {K.}~\bibnamefont {Luke}}, \bibinfo {author}
  {\bibfnamefont {E.~S.}\ \bibnamefont {Lamb}}, \bibinfo {author}
  {\bibfnamefont {M.~R.}\ \bibnamefont {Lamont}}, \bibinfo {author}
  {\bibfnamefont {C.}~\bibnamefont {Joshi}}, \bibinfo {author} {\bibfnamefont
  {Y.}~\bibnamefont {Okawachi}}, \bibinfo {author} {\bibfnamefont {F.~W.}\
  \bibnamefont {Wise}}, \bibinfo {author} {\bibfnamefont {M.}~\bibnamefont
  {Lipson}}, \ and\ \bibinfo {author} {\bibfnamefont {A.}~\bibnamefont
  {Gaeta}},\ }\href {\doibase 10.1364/OL.40.005117} {\bibfield  {journal}
  {\bibinfo  {journal} {Optics letters}\ }\textbf {\bibinfo {volume} {40}},\
  \bibinfo {pages} {5117} (\bibinfo {year} {2015})}\BibitemShut {NoStop}%
\bibitem [{\citenamefont {Guo}\ \emph {et~al.}(2018)\citenamefont {Guo},
  \citenamefont {Herkommer}, \citenamefont {Billat}, \citenamefont {Grassani},
  \citenamefont {Zhang}, \citenamefont {Pfeiffer}, \citenamefont {Weng},
  \citenamefont {Br{\`e}s},\ and\ \citenamefont {Kippenberg}}]{guo2018mid}%
  \BibitemOpen
  \bibfield  {author} {\bibinfo {author} {\bibfnamefont {H.}~\bibnamefont
  {Guo}}, \bibinfo {author} {\bibfnamefont {C.}~\bibnamefont {Herkommer}},
  \bibinfo {author} {\bibfnamefont {A.}~\bibnamefont {Billat}}, \bibinfo
  {author} {\bibfnamefont {D.}~\bibnamefont {Grassani}}, \bibinfo {author}
  {\bibfnamefont {C.}~\bibnamefont {Zhang}}, \bibinfo {author} {\bibfnamefont
  {M.~H.}\ \bibnamefont {Pfeiffer}}, \bibinfo {author} {\bibfnamefont
  {W.}~\bibnamefont {Weng}}, \bibinfo {author} {\bibfnamefont {C.-S.}\
  \bibnamefont {Br{\`e}s}}, \ and\ \bibinfo {author} {\bibfnamefont {T.~J.}\
  \bibnamefont {Kippenberg}},\ }\href {\doibase 10.1038/s41566-018-0144-1}
  {\bibfield  {journal} {\bibinfo  {journal} {Nature Photonics}\ }\textbf
  {\bibinfo {volume} {12}},\ \bibinfo {pages} {330} (\bibinfo {year}
  {2018})}\BibitemShut {NoStop}%
\bibitem [{\citenamefont {Pfeiffer}\ \emph {et~al.}(2018)\citenamefont
  {Pfeiffer}, \citenamefont {Herkommer}, \citenamefont {Liu}, \citenamefont
  {Morais}, \citenamefont {Zervas}, \citenamefont {Geiselmann},\ and\
  \citenamefont {Kippenberg}}]{pfeiffer2018photonic}%
  \BibitemOpen
  \bibfield  {author} {\bibinfo {author} {\bibfnamefont {M.~H.~P.}\
  \bibnamefont {Pfeiffer}}, \bibinfo {author} {\bibfnamefont {C.}~\bibnamefont
  {Herkommer}}, \bibinfo {author} {\bibfnamefont {J.}~\bibnamefont {Liu}},
  \bibinfo {author} {\bibfnamefont {T.}~\bibnamefont {Morais}}, \bibinfo
  {author} {\bibfnamefont {M.}~\bibnamefont {Zervas}}, \bibinfo {author}
  {\bibfnamefont {M.}~\bibnamefont {Geiselmann}}, \ and\ \bibinfo {author}
  {\bibfnamefont {T.~J.}\ \bibnamefont {Kippenberg}},\ }\href {\doibase
  10.1109/JSTQE.2018.2808258} {\bibfield  {journal} {\bibinfo  {journal} {IEEE
  Journal of Selected Topics in Quantum Electronics}\ }\textbf {\bibinfo
  {volume} {24}},\ \bibinfo {pages} {1} (\bibinfo {year} {2018})}\BibitemShut
  {NoStop}%
\bibitem [{\citenamefont {Liu}\ \emph {et~al.}(2018)\citenamefont {Liu},
  \citenamefont {Raja}, \citenamefont {Karpov}, \citenamefont {Ghadiani},
  \citenamefont {Pfeiffer}, \citenamefont {Du}, \citenamefont {Engelsen},
  \citenamefont {Guo}, \citenamefont {Zervas},\ and\ \citenamefont
  {Kippenberg}}]{Liu:18}%
  \BibitemOpen
  \bibfield  {author} {\bibinfo {author} {\bibfnamefont {J.}~\bibnamefont
  {Liu}}, \bibinfo {author} {\bibfnamefont {A.~S.}\ \bibnamefont {Raja}},
  \bibinfo {author} {\bibfnamefont {M.}~\bibnamefont {Karpov}}, \bibinfo
  {author} {\bibfnamefont {B.}~\bibnamefont {Ghadiani}}, \bibinfo {author}
  {\bibfnamefont {M.~H.~P.}\ \bibnamefont {Pfeiffer}}, \bibinfo {author}
  {\bibfnamefont {B.}~\bibnamefont {Du}}, \bibinfo {author} {\bibfnamefont
  {N.~J.}\ \bibnamefont {Engelsen}}, \bibinfo {author} {\bibfnamefont
  {H.}~\bibnamefont {Guo}}, \bibinfo {author} {\bibfnamefont {M.}~\bibnamefont
  {Zervas}}, \ and\ \bibinfo {author} {\bibfnamefont {T.~J.}\ \bibnamefont
  {Kippenberg}},\ }\href {\doibase 10.1364/OPTICA.5.001347} {\bibfield
  {journal} {\bibinfo  {journal} {Optica}\ }\textbf {\bibinfo {volume} {5}},\
  \bibinfo {pages} {1347} (\bibinfo {year} {2018})}\BibitemShut {NoStop}%
\bibitem [{\citenamefont {Drake}\ \emph {et~al.}(2018)\citenamefont {Drake},
  \citenamefont {Stone}, \citenamefont {Briles}, \citenamefont {Spencer},\ and\
  \citenamefont {Papp}}]{drake2018thermal}%
  \BibitemOpen
  \bibfield  {author} {\bibinfo {author} {\bibfnamefont {T.~E.}\ \bibnamefont
  {Drake}}, \bibinfo {author} {\bibfnamefont {J.~R.}\ \bibnamefont {Stone}},
  \bibinfo {author} {\bibfnamefont {T.~C.}\ \bibnamefont {Briles}}, \bibinfo
  {author} {\bibfnamefont {D.~T.}\ \bibnamefont {Spencer}}, \ and\ \bibinfo
  {author} {\bibfnamefont {S.~B.}\ \bibnamefont {Papp}},\ }in\ \href {\doibase
  10.1364/FIO.2018.JTu2A.62} {\emph {\bibinfo {booktitle} {Frontiers in
  Optics}}}\ (\bibinfo {organization} {Optical Society of America},\ \bibinfo
  {year} {2018})\ pp.\ \bibinfo {pages} {JTu2A--62}\BibitemShut {NoStop}%
\bibitem [{\citenamefont {Arbabi}\ and\ \citenamefont {Goddard}(2013)}]{dndt}%
  \BibitemOpen
  \bibfield  {author} {\bibinfo {author} {\bibfnamefont {A.}~\bibnamefont
  {Arbabi}}\ and\ \bibinfo {author} {\bibfnamefont {L.~L.}\ \bibnamefont
  {Goddard}},\ }\href {\doibase 10.1364/OL.38.003878} {\bibfield  {journal}
  {\bibinfo  {journal} {Opt. Lett.}\ }\textbf {\bibinfo {volume} {38}},\
  \bibinfo {pages} {3878} (\bibinfo {year} {2013})}\BibitemShut {NoStop}%
\bibitem [{\citenamefont {Lim}\ \emph {et~al.}(2017)\citenamefont {Lim},
  \citenamefont {Savchenkov}, \citenamefont {Dale}, \citenamefont {Liang},
  \citenamefont {Eliyahu}, \citenamefont {Ilchenko}, \citenamefont {Matsko},
  \citenamefont {Maleki},\ and\ \citenamefont {Wong}}]{Lim2017a}%
  \BibitemOpen
  \bibfield  {author} {\bibinfo {author} {\bibfnamefont {J.}~\bibnamefont
  {Lim}}, \bibinfo {author} {\bibfnamefont {A.~A.}\ \bibnamefont {Savchenkov}},
  \bibinfo {author} {\bibfnamefont {E.}~\bibnamefont {Dale}}, \bibinfo {author}
  {\bibfnamefont {W.}~\bibnamefont {Liang}}, \bibinfo {author} {\bibfnamefont
  {D.}~\bibnamefont {Eliyahu}}, \bibinfo {author} {\bibfnamefont
  {V.}~\bibnamefont {Ilchenko}}, \bibinfo {author} {\bibfnamefont {A.~B.}\
  \bibnamefont {Matsko}}, \bibinfo {author} {\bibfnamefont {L.}~\bibnamefont
  {Maleki}}, \ and\ \bibinfo {author} {\bibfnamefont {C.~W.}\ \bibnamefont
  {Wong}},\ }\href {\doibase 10.1038/s41467-017-00021-9} {\bibfield  {journal}
  {\bibinfo  {journal} {Nature Communications}\ }\textbf {\bibinfo {volume}
  {8}},\ \bibinfo {pages} {8} (\bibinfo {year} {2017})}\BibitemShut {NoStop}%
\bibitem [{\citenamefont {Drever}\ \emph {et~al.}(1983)\citenamefont {Drever},
  \citenamefont {Hall}, \citenamefont {Kowalski}, \citenamefont {Hough},
  \citenamefont {Ford}, \citenamefont {Munley},\ and\ \citenamefont
  {Ward}}]{drever1983laser}%
  \BibitemOpen
  \bibfield  {author} {\bibinfo {author} {\bibfnamefont {R.}~\bibnamefont
  {Drever}}, \bibinfo {author} {\bibfnamefont {J.~L.}\ \bibnamefont {Hall}},
  \bibinfo {author} {\bibfnamefont {F.}~\bibnamefont {Kowalski}}, \bibinfo
  {author} {\bibfnamefont {J.}~\bibnamefont {Hough}}, \bibinfo {author}
  {\bibfnamefont {G.}~\bibnamefont {Ford}}, \bibinfo {author} {\bibfnamefont
  {A.}~\bibnamefont {Munley}}, \ and\ \bibinfo {author} {\bibfnamefont
  {H.}~\bibnamefont {Ward}},\ }\href {\doibase 10.1007/BF00702605} {\bibfield
  {journal} {\bibinfo  {journal} {Applied Physics B}\ }\textbf {\bibinfo
  {volume} {31}},\ \bibinfo {pages} {97} (\bibinfo {year} {1983})}\BibitemShut
  {NoStop}%
\bibitem [{\citenamefont {Kimble}\ \emph {et~al.}(2008)\citenamefont {Kimble},
  \citenamefont {Lev},\ and\ \citenamefont {Ye}}]{noisereduction}%
  \BibitemOpen
  \bibfield  {author} {\bibinfo {author} {\bibfnamefont {H.~J.}\ \bibnamefont
  {Kimble}}, \bibinfo {author} {\bibfnamefont {B.~L.}\ \bibnamefont {Lev}}, \
  and\ \bibinfo {author} {\bibfnamefont {J.}~\bibnamefont {Ye}},\ }\href
  {\doibase 10.1103/PhysRevLett.101.260602} {\bibfield  {journal} {\bibinfo
  {journal} {Phys. Rev. Lett.}\ }\textbf {\bibinfo {volume} {101}},\ \bibinfo
  {pages} {260602} (\bibinfo {year} {2008})}\BibitemShut {NoStop}%
\end{thebibliography}%

\end{document}